# Femtosecond-Nanometer Visualization of Exciton Dynamics in MoSe$_2$

Paul M. Sass and Patrick Z. El-Khoury*

*Physical Sciences Division, Pacific Northwest National Laboratory, P.O. Box 999, Richland, Washington 99352, United States*

*Correspondence to: patrick.elKhoury@pnnl.gov (PZE)

**ABSTRACT**

Femtosecond (fs)-resolved photoemission electron micrographs of single and few-layer $MoSe_2$ track exciton dynamics in this model 2D quantum material system with joint nanometer (nm)-fs resolution. We illustrate the latter using two-photon photoemission following 400 nm fs irradiation to probe and tunable near-IR excitation (770-840 nm) to populate the lowest accessible A-exciton states in $MoSe_2$. Through distinct imaging contrasts in single/few-layer regions, we spatio-temporally visualize exciton lifetimes in multilayers that are ~24 times longer than their monolayer analogs. We also find that tuning the near-IR excitation source allows us to track exciton resonances that vary on the nm scale. In effect, our approach allows us to track exciton resonances and dynamics in real-space and in real-time, which is not currently possible (or at least trivially implemented) using conventional all-optical, nano-optical, and all-electron-based microscopies.

**TOC Graphic**

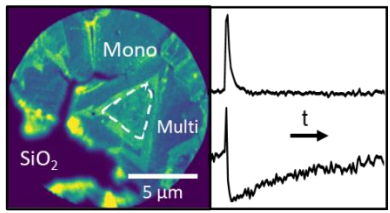

Two-dimensional semiconducting transition metal dichalcogenides (TMDs) have been popularized because of their exotic physical properties that arise from their unique band structures and reduced dimensionalities.[1] Properties such as a direct bandgap in monolayers,[2-3] valley polarization,[4] and the existence of bandgaps 'on-demand' because of the extensive library of TMDs[5] all come to mind in this context. The drive to understand these systems is not only bolstered by their emerging applications in (micro-)electronics, but it also arises from their unique vibrational, electronic, and optical properties more generally.[6-8] Indeed, TMDs have found applications in light-harvesting,[9] ultrasensitive photodetectors,[10] field-effect transistors,[11] and light-emitting diodes.[12] At the core of the intriguing properties of single/few-layer TMDs are strongly bound excitons that are observable at room temperature and that feature binding energies on the order of hundreds of meV's.[13-16] In this context, the efficiencies of TMDs are somewhat limited by exciton diffusion and recombination rates; lower efficiencies in (proto)devices can be associated with non-radiative recombination and/or trapping of electrons and holes at (often nanoscopic) defect sites.[17-18]

Ultrafast spectroscopy and micro-spectroscopy are powerful tools that may be used to track exciton dynamics in TMDs, as noted in related prior works.[19-22] Such tools have revealed stark variances in exciton lifetimes in monolayer and multilayer TMDs. Radiative lifetimes can be on the order of 100 ps to 1 ns, and even 11 ns with minimal pump fluences and surface passivation.[23] The latter is perhaps indicative of non-radiative/Auger-type processes, possibly at defects, which obscure the intrinsic exciton radiative lifetimes. This in part motivates our quest towards femtosecond nanoscopy of exciton dynamics in 2D quantum materials. Additionally, differences in dynamics between samples with a different number of layers is not entirely understood. In one study, exciton dynamics of exfoliated $MoS_2$ layers revealed intravalley relaxation rates that can be enhanced 40-fold in monolayers compared to thicker samples.[24] Nevertheless, layer dependent properties deserve more systematic exploration, particularly since most all-optical setups are diffraction-limited, and hence, cannot be used to track nanoscale variations in exciton lifetimes and resonances. The latter may be key to understanding the sample to sample-dependent properties that are well-documented in literature.

Herein, we employ transient photoemission electron microscopy (tr-PEEM) to visualize exciton dynamics in monolayer and multilayer $MoSe_2$. Our approach generally affords femtosecond temporal and nanometer spatial resolution. For a more detailed discussion of time-resolved PEEM measurements, the reader is referred to previous work from our group and others.[25-28] In this study, an ultrashort pulse (~25 fs) is tuned to coincide with the first excitonic transition (~810 nm, ~1.53 eV) of $MoSe_2$. Towards the end of this letter, we also show that we can limit the bandwidth of our near-IR excitation source and tune it over the exciton resonance to track the exciton energy through photoemission. A second pulse (405 nm, 120 fs) induces two-photon photoemission both from monolayer- and multi-layer regions across the field of view (see Supporting Information for non-linear power dependence). By changing the relative time delay between the red and blue pulses, the PEEM images track exciton dynamics through disparate contrasts (see below) from monolayers and multilayers of $MoSe_2$.

Prior to tr-PEEM mapping, combined atomic force microscopy (AFM) and tip-enhanced photoluminescence (TEPL) measurements were used to distinguish between monolayers and multilayers across our sample. Figure 1 shows AFM and TEPL images of CVD-grown $MoSe_2$ triangles that have been exfoliated onto $SiO_2$. The larger area scan in Figure 1a shows bright multilayered triangles surrounded by non-uniformly distributed monolayer areas, marked by measurable TEPL signals. The brightest region running across the image is an area of thicker

MoSe$_2$ ($\geq$ ~10 layers). Figure 1b shows an AFM image of the dashed red square in Figure 1a, which features multilayered MoSe$_2$ triangles. The number of layers is ~2-6 as shown in the topographic line cut (along the red-dotted line) in Figure 1d. The surrounding area consists of monolayers, as confirmed by TEPL mapping in Figure 1c. A spatially averaged spectrum of the TEPL (red-dashed circle) is plotted in Figure 1d and exhibits a clear resonance centered at ~785 nm. This, coupled with the topography, establishes multilayer *vs* monolayer areas across our sample.

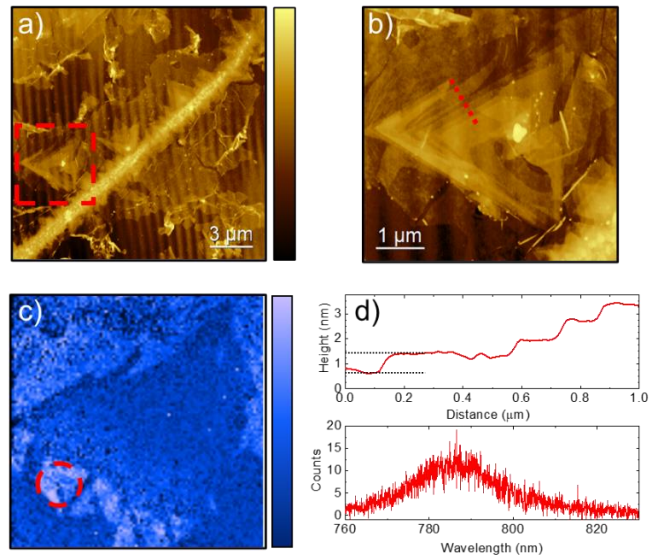

Figure 1. (a) Topographic image of exfoliated CVD-grown MoSe$_2$ triangles on SiO$_2$ showing a mix of monolayer and multilayer (2-10) regions. (b) Enhanced topographic image of the region highlighted in (a). (c) TEPL near-field image of the area shown in (b), revealing contrast arising from TEPL from monolayers. (d) Top: Topographic line-cut along the red dotted line in (b) showing disparate steps. Bottom: Spatially averaged TEPL spectrum taken from the circular area highlighted in (c). The scale bars for the topographic and TEPL images range from 0 to 26 nm (a), 10 nm (b), and 6000 counts (c).

We next turn to tr-PEEM mapping to explore the distinct spatiotemporal dynamics of excitons in monolayer and multilayer regions identified through combined AFM-TEPL mapping. The data consists of a series of PEEM images recorded as a function of time delay between the pump (810 nm) and probe (405 nm) pulses. In other words, each pixel in any given image contains information about the dynamics in that region of the spectrum. Schematics of two assignable nonlinear photoemission processes encountered/discussed in the ensuing sections of this work are shown in Figure 2. In both schemes (termed *a* and *b* in Figure 2), the red pulse is tuned into resonance with the first excitonic state, and the blue pulse is used to monitor the dynamics of the excited excitons through nonlinear (two-photon) photoemission. Blue pulse-induced nonlinear photoemission can take place both from the ground (a) and excited excitonic state (b). Both processes occur simultaneously, and photoemission cross-sections from the ground *vs* lowest excitonic state govern the relative contributions of processes *a* and *b* to the overall signals. It is important to stress that the dynamics of excited excitons are tracked in both cases. Namely, either ground state population and decay (process *b*) or depopulation and recovery (process *a*) govern the recorded spatiotemporally resolved PEEM maps.

Both the transient images as well as the dynamic traces that are encoded in each pixel can be used to describe our tr-PEEM results. We start with the images. In time-resolved measurements, relatively bright two-photon photoemission from the multilayer regions was observed at negative time, i.e., prior to the arrival of the red pulse, relative to monolayer regions, see Figure 3a. At time zero, when the red and blue pulses are time-coincident, we see a rise in photoemission from the monolayer regions, accompanied by decreased photoemission from the multilayers, see Figure 3b. At positive time delay, we observe a rapid decay of the photoemission intensity from the monolayers and a slower recovery of the diminished photoemission signal from the multilayer regions. Based on these observations, the dynamics we observe are consistent with exciton state population and decay (process *b* in Figure 2) for the monolayers and ground state de-population and recovery (process *a* in Figure 2) for the multilayers. The measured decay times and near-IR excitation wavelength dependence that we discuss next further corroborate these assignments.

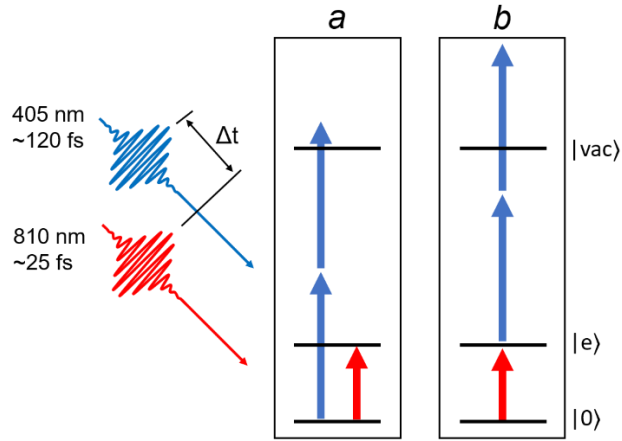

Figure 2. Schematics of two processes that contribute to tr-PEEM maps of monolayer and multilayer MoSe$_2$. See text for more details.

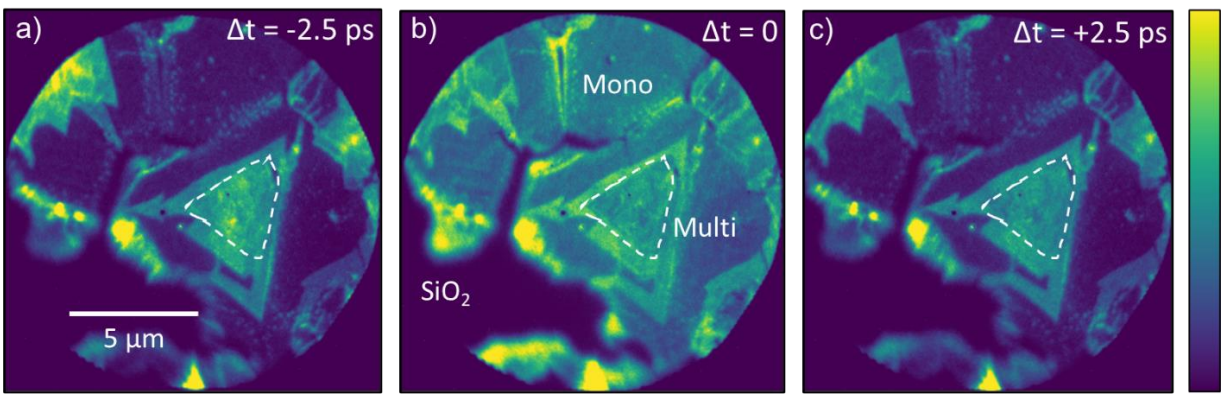

Figure 3. (a-c) Tr-PEEM images of a small region of interest on our sample, showing varying time-dependent contrasts at regions corresponding to monolayer and multilayer MoSe$_2$. The images were taken at negative time (a, $\Delta t$ = -2.5 ps), time zero (b, $\Delta t$ = 0 ps), and positive delay time (c, $\Delta t$ = +2.5 ps). The white dashed irregular triangular area traced in all three panels highlights a multilayer region of the sample, with surrounding monolayer and bilayers. The monolayer areas become bright only at time zero, while the multilayer area exhibits a decrease in photoemission at time zero and signal recovery at positive time.

As mentioned above, each pixel in the images shown in Figure 3 contains dynamic information. Spatially averaged tr-PEEM signals can thus be used to examine the temporal dynamics that are encoded in the monolayer and multilayer regions more closely. The results are shown in Figure 4: long time and short time dynamics are plotted in panels a and b, respectively.

As alluded to above, transient photoemission signals in monolayer regions feature a spike at time zero, followed by an exponential decay, which can be attributed to the excitation and decay of the excited excitons. In contrast, the multilayers first appear relatively bright at negative time. A dramatic drop in photoemission occurs when the red and blue lasers are time coincident. At positive time, the photoemission signal slowly recovers to its original level. Using the two distinct processes, we measure exciton lifetimes of 9.38/0.4 ps in multilayers/monolayers, respectively. These values are generally consistent with literature values for this and structurally related 2D quantum materials systems. [21-22]

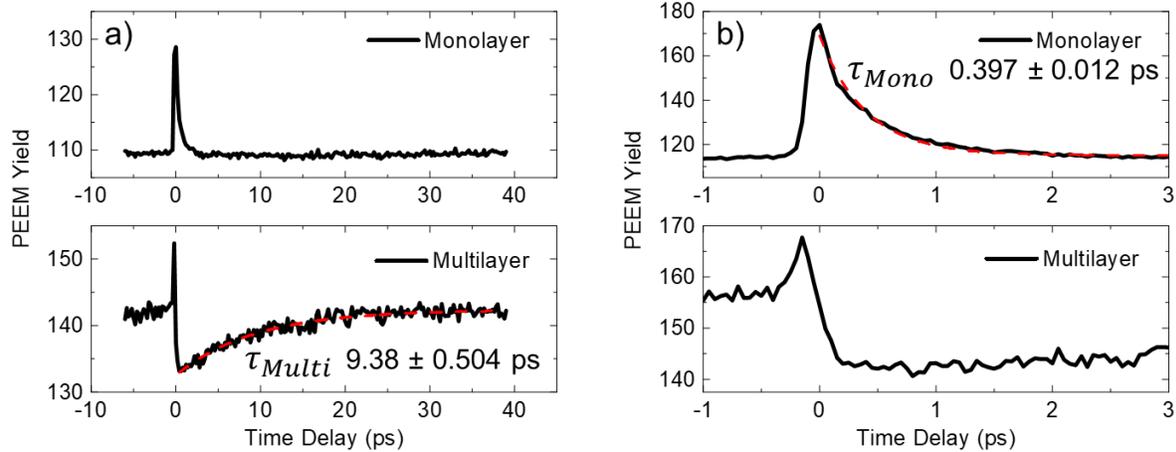

Figure 4. (a,b) Spatially averaged tr-PEEM signals on a monolayer area (top) and multilayer area (bottom). (b) Spatially averaged PEEM signals with higher time resolution. The monolayer shows photoemission decay from the exciton state. The multilayer, however, shows decreased photoemission at time zero and recovery as the exciton state decays. Red-dotted lines show exponential decay fitting

These results are best understood in the context of recent PEEM works on related materials systems. For instance, Gong and co-workers examined supported and suspended $WS_2$ monolayers using time and energy-resolved PEEM.[21] The authors resolved electron cooling in the Q valley of the conduction band and associated electron decay with defect trapping. Another more recent study by Yang and co-workers[22] combined several techniques, including transient PEEM and photoelectron spectroscopy to understand the role of defects on the exciton dynamics in $MoS_2$ flakes. It was found that defects serve as electron-hole recombination cites that greatly reduce carrier lifetimes. Overall, relatively fast recombination rates like the ones we observe are associated with exciton trapping at defect sites, whereas slower processes (0.1-1 nanoseconds) are generally attributed to radiative exciton recombination. On the basis of the prior observations, it is possible to associate the observed fast dynamics from both the monolayer and multilayer regions with electron-hole recombination at defect sites. Note that the finite micrometer sized nanoflakes we imaged in this work also complicate the interpretation of the dynamics, since the edges can play a significant role in exciton recombination.

We also varied the central wavelength of the red pulse over the first exciton resonance to test whether excitation energy tunable tr-PEEM imaging may be used to track not just the dynamics, but also the resonances of the excitons with nanometer spatial resolution. In both the monolayer and multilayer traces, the amplitudes of the spikes at time zero, like that in Figure 4, appear to be highest at the center of the exciton resonance, and lowest towards the blue and red wings of the same resonance. However, the raw amplitude at time zero can depend on a few different variables, such as input power and phase interference effects. By looking at the relative difference in photoemission before, at, and after time zero, these effects are avoided. For the monolayer, the difference in photoemission at time zero and negative time is shown by the blue curve in Figure 5. For the multilayer, the difference between the signals averaged from -0.65 ps to -0.35 ps and 0.4 ps to 1.25 ps traces the red curve in Figure 5 (see Supporting Information for full kinetic traces). The two curves agree well with the spatially averaged spectrum (grey curve in Figure 5) that was taken from a hyperspectral optical absorption microscopy image recorded from a different multilayer $MoSe_2$ sample (see Supporting Information). Indeed, the monolayer and multilayer PEEM-derived resonances peak around 800 and 805 nm, respectively. A closer inspection of the blue and red traces nonetheless reveals that multiple sharp peaks contribute to the blue and red traces. These peaks sharper peaks vary in relative intensities at different spatial positions (see Supporting Information). Given the spatial resolution of our measurements (20-50 nm), we associate these bands to spatially varying exciton resonances/to the removal of averaging that mars diffraction limited optical and photoelectron measurements. The latter will be more thoroughly addressed in follow-up studies.

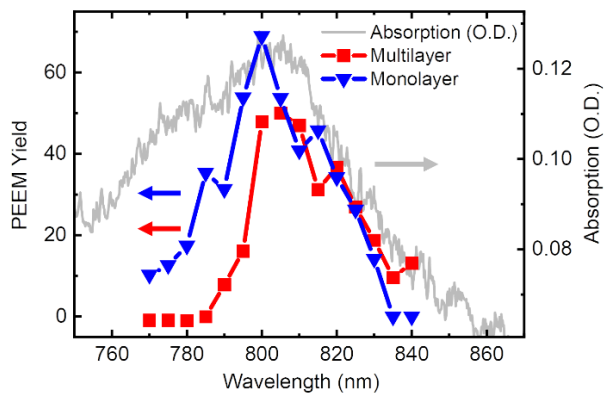

Figure 5. Tr-PEEM signal contrast obtained from monolayer (blue) and multilayer (red) areas as a function of red laser excitation wavelength. The multilayer signal is calculated as the difference between signals averaged from -0.65 ps to -0.35 ps and from 0.4 ps to 1.25 ps. The monolayer signal consists of the difference between the peak intensity at time zero and the signal at negative time. The grey curve traces the A-exciton resonance using through and through optical absorption. The sample used to measure absorption consists of $MoSe_2$ exfoliated on glass.

In summary, nonlinear PEEM mapping was used for spectrally, spatially, and temporally-resolved characterization of exciton dynamics in monolayer and multilayer $MoSe_2$. Two distinct PEEM transient imaging contrast mechanisms allowed us to visualize exciton dynamics in monolayers and multilayers. The measured spatially-resolved exciton lifetime in the multilayers was found to be ~24 times longer than its analog in monolayers. Finally, tuning the wavelength of the incident red pulse traces the exciton resonance, as confirmed through separate hyperspectral optical absorption measurements. Overall, the measurements reported here and their analysis should apply to a broad range of 2D quantum materials. Indeed, the ability to resolve exciton dynamics and resonances in real space may hold the key to understanding how nanoscopic defects (native or engineered) affect the optical and electronic properties of TMDs and TMD-featuring devices.

**METHODS**

The sample consists of CVD grown $MoSe_2$ triangles that were been exfoliated onto $SiO_2$/Si. It was purchased from 2Dsemiconductors USA and used without further modification.

AFM and TEPL measurements were performed using a previously described optical setup[29]. The incident laser power was on the order of a few 100's of µWs, focused onto the apex of an Au tip using a 100X long working distance air objective (NA=0.7). The polarization of the incident laser was along the long axis of the probe, which was prepared by sputtering a conventional silicon probe (Nanosensors, A) with 100 nm of gold.

Our PEEM setup is described elsewhere in more detail.[28, 30] Briefly, a mode-locked Ti:sapphire laser produces 15 fs pulses centered at 810 nm, at a repetition rate of 90 MHz. Approximately 80% of the pulse is split to produce the second harmonic in a 500 µm thick BBO crystal. Group velocity mismatch in the second harmonic crystal leads to blue pulses with pulse durations near 90 fs. A variable delay line controls the relative timing between the red (~810 nm) and blue (~410 nm) pulses. *P*-polarized laser pulses are recombined on a dichroic beam splitter and collinearly directed onto the PEEM sample at a 75° angle of incidence with respect to the surface normal. The spot sizes of the separate beams are adjusted such that typically the red pulse spot size is roughly 50% smaller than the blue pulse spot size at the sample position. Power dependences are determined by measuring the photoelectron yield as a function of incident laser power. Two-color power dependence experiments were performed by holding one pulse at a constant power while varying the power of the second pulse. These results are shown in the supporting information section of our work.

## SUPPORTING INFORMATION

Power-dependence of PEEM signal, hyperspectral optical absorption mapping, wavelength-dependent tr-PEEM kinetics , and exciton resonance spatial dependence.

## ACKNOWLEDGMENTS

PMS and PZE acknowledge support from the United States Department of Energy, Office of Science, Office of Basic Energy Sciences, Division of Chemical Sciences, Geosciences & Biosciences. Some of the instrumentation that was required to do the measurements described in this work was purchased and developed through funding from the Laboratory Directed Research and Development (LDRD) program at Pacific Northwest National Laboratory. The authors are grateful to Alan Joly and Kevin Crampton (PNNL) for valuable discussions and to Chih-Feng Wang (PNNL) for the hyperspectral optical measurements.

## AUTHOR INFORMATION

**Corresponding Author**

*patrick.elkhoury@pnnl.gov

The authors declare no competing financial interest.

## REFERENCES

1. Novoselov, K. S.; Jiang, D.; Schedin, F.; Booth, T. J.; Khotkevich, V. V.; Morozov, S. V.; Geim, A. K. Two-dimensional atomic crystals. *Proc. Natl. Acad. Sci. U.S.A* **2005,** *102* , 10451-10453.
2. Splendiani, A.; Sun, L.; Zhang, Y.; Li, T.; Kim, J.; Chim, C.-Y.; Galli, G.; Wang, F. Emerging photoluminescence in monolayer $MoS_2$. *Nano Lett.* **2010,** *10* , 1271-1275.
3. Mak, K. F.; Lee, C.; Hone, J.; Shan, J.; Heinz, T. F. Atomically thin $MoS_2$: A new direct-gap semiconductor. *Phys. Rev.Lett.* **2010,** *105* , 136805.
4. Zeng, H.; Dai, J.; Yao, W.; Xiao, D.; Cui, X. Valley polarization in MoS2 monolayers by optical pumping. *Nature Nanotechnol.* **2012,** *7* , 490-493.
5. Duan, X.; Wang, C.; Pan, A.; Yu, R.; Duan, X. Two-dimensional transition metal dichalcogenides as atomically thin semiconductors: opportunities and challenges. *Chem. Soc. Rev.* **2015,** *44* , 8859-8876.
6. Srivastava, A.; Sidler, M.; Allain, A. V.; Lembke, D. S.; Kis, A.; Imamoğlu, A. Valley Zeeman effect in elementary optical excitations of monolayer $WSe_2$. *Nature Phys.* **2015,** *11* , 141-147.
7. Berkelbach, T. C.; Reichman, D. R. Optical and excitonic properties of atomically thin transition-metal dichalcogenides. *Ann. Rev. Cond. Matt. Phys.* **2018,** *9* , 379-396.
8. Mueller, T.; Malic, E. Exciton physics and device application of two-dimensional transition metal dichalcogenide semiconductors. *NPJ 2D Materials and Applications* **2018,** *2* , 29.
9. Bernardi, M.; Palummo, M.; Grossman, J. C. Extraordinary sunlight absorption and one nanometer thick photovoltaics using two-dimensional monolayer materials. *Nano Lett.* **2013,** *13*, 3664-3670.
10. Baugher, B. W.; Churchill, H. O.; Yang, Y.; Jarillo-Herrero, P. Optoelectronic devices based on electrically tunable p-n diodes in a monolayer dichalcogenide. *Nat. Nanotechnol.* **2014,** *9*, 262-267.
11. Yuan, H.; Bahramy, M. S.; Morimoto, K.; Wu, S.; Nomura, K.; Yang, B.-J.; Shimotani, H.; Suzuki, R.; Toh, M.; Kloc, C.; Xu, X.; Arita, R.; Nagaosa, N.; Iwasa, Y. Zeeman-type spin splitting controlled by an electric field. *Nature Phys.* **2013,** *9*, 563-569.
12. Pospischil, A.; Furchi, M. M.; Mueller, T. Solar-energy conversion and light emission in an atomic monolayer p-n diode. *Nat. Nanotechnol.* **2014,** *9*, 257-261.
13. Qiu, D. Y.; da Jornada, F. H.; Louie, S. G. Optical spectrum of $MoS_2$: many-body effects and diversity of exciton states. *Phys. Rev. Lett.* **2013,** *111* , 216805.
14. Molina-Sánchez, A.; Sangalli, D.; Hummer, K.; Marini, A.; Wirtz, L. Effect of spin-orbit interaction on the optical spectra of single-layer, double-layer, and bulk $MoS_2$. *Phys. Rev. B* **2013,** *88*, 045412.
15. Park, S.; Mutz, N.; Schultz, T.; Blumstengel, S.; Han, A.; Aljarb, A.; Li, L.-J.; List-Kratochvil, E. J. W.; Amsalem, P.; Koch, N. Direct determination of monolayer $MoS_2$ and $WSe_2$ exciton binding energies on insulating and metallic substrates. *2D Materials* **2018,** *5*, 025003.
16. Hill, H. M.; Rigosi, A. F.; Roquelet, C.; Chernikov, A.; Berkelbach, T. C.; Reichman, D. R.; Hybertsen, M. S.; Brus, L. E.; Heinz, T. F. Observation of excitonic rydberg states in monolayer $MoS_2$ and $WS_2$ by photoluminescence excitation spectroscopy. *Nano Lett.* **2015,** *15*, 2992-2997.
17. Poellmann, C.; Steinleitner, P.; Leierseder, U.; Nagler, P.; Plechinger, G.; Porer, M.; Bratschitsch, R.; Schüller, C.; Korn, T.; Huber, R. Resonant internal quantum transitions and femtosecond radiative decay of excitons in monolayer $WSe_2$. *Nat. Mater.* **2015,** *14*, 889-893.
18. Sun, D.; Rao, Y.; Reider, G. A.; Chen, G.; You, Y.; Brézin, L.; Harutyunyan, A. R.; Heinz, T. F. Observation of rapid exciton–exciton annihilation in monolayer molybdenum disulfide. *Nano Lett.* **2014,** *14* , 5625-5629.
19. Wang, L.; Xu, C.; Li, M.-Y.; Li, L.-J.; Loh, Z.-H. Unraveling spatially heterogeneous ultrafast carrier dynamics of single-layer $WSe_2$ by femtosecond time-resolved photoemission electron microscopy. *Nano Lett.* **2018,** *18*, 5172-5178.
20. Madeo, J.; Man, M. K. L.; Sahoo, C.; Campbell, M.; Pareek, V.; Wong, E. L.; Al-Mahboob, A.; Chan, N. S.; Karmakar, A.; Mariserla, B. M. K.; Li, X. Q.; Heinz, T. F.; Cao, T.; Dani, K. M. Directly visualizing the momentum-forbidden dark excitons and their dynamics in atomically thin semiconductors. *Science* **2020,** *370* , 1199-1203.
21. Li, Y. L.; Liu, W.; Wang, Y. K.; Xue, Z. H.; Leng, Y. C.; Hu, A. Q.; Yang, H.; Tan, P. H.; Liu, Y. Q.; Misawa, H.; Sun, Q.; Gao, Y. N.; Hu, X. Y.; Gong, Q. H. Ultrafast electron cooling and decay in monolayer WS$_2$ revealed by time- and energy-resolved photoemission electron microscopy. *Nano Lett.* **2020,** *20* , 3747-3753.
22. Liang, Y.; Li, B.-H.; Li, Z.; Zhang, G.; Sun, J.; Zhou, C.; Tao, Y.; Ye, Y.; Ren, Z.; Yang, X. Spatially heterogeneous ultrafast interfacial carrier dynamics of 2D-MoS2 flakes. *Materials Today Physics* **2021,** *21*, 100506.
23. Moody, G.; Schaibley, J.; Xu, X. Exciton dynamics in monolayer transition metal dichalcogenides. *J. Opt. Soc. Am. B* **2016,** *33* , C39-c49.
24. Shi, H.; Yan, R.; Bertolazzi, S.; Brivio, J.; Gao, B.; Kis, A.; Jena, D.; Xing, H. G.; Huang, L. Exciton dynamics in suspended monolayer and few-layer MoS$_2$ 2D crystals. *ACS Nano* **2013,** *7* , 1072-1080.
25. Stockman, M. I.; Kling, M. F.; Kleineberg, U.; Krausz, F. Attosecond nanoplasmonic-field microscope. *Nature Photonics* **2007,** *1* , 539-544.
26. Kubo, A.; Onda, K.; Petek, H.; Sun, Z.; Jung, Y. S.; Kim, H. K. Femtosecond imaging of surface plasmon dynamics in a nanostructured silver film. *Nano Lett.* **2005,** *5* , 1123-1127.
27. Kubo, A.; Pontius, N.; Petek, H. Femtosecond microscopy of surface plasmon polariton wave packet evolution at the silver/vacuum interface. *Nano Lett.* **2007,** *7* , 470-475.
28. Gong, Y.; Joly, A. G.; Hu, D.; El-Khoury, P. Z.; Hess, W. P. Ultrafast imaging of surface plasmons propagating on a dold surface. *Nano Lett.* **2015,** *15* , 3472-3478.
29. Wang, C.-F.; O'Callahan, B. T.; Krayev, A.; El-Khoury, P. Z. Nanoindentation-enhanced tip-enhanced Raman spectroscopy. *J. Chem. Phys.* **2021,** *154* , 241101.
30. Joly, A. G.; El-Khoury, P. Z.; Hess, W. P. Spatiotemporal imaging of surface plasmons using two-color photoemission electron microscopy. *J. Phys. Chem. C* **2018,** *122* , 20981-20988.